\documentclass[12pt,onecolumn,oneside]{IEEEtran}
\usepackage{multirow,hhline}

\usepackage{enumerate}

\usepackage{amsbsy}
\usepackage{amssymb}
\usepackage{amscd}
\usepackage{amsmath}
\usepackage{graphics}
\usepackage{graphicx}
\usepackage{epstopdf}
\usepackage{psfrag}
\def\bPsi{{\mathbf{\Psi}}}

\def\bx{{\mathbf{x}}}
\def\by{{\mathbf{y}}}
\def\bz{{\mathbf{z}}}
\def\bA{{\mathbf{A}}}

\def\bC{{\mathbf{C}}}
\def\bD{{\mathbf{D}}}
\def\bE{{\mathbf{E}}}
\def\bF{{\mathbf{F}}}

\def\bH{{\mathbf{H}}}

\def\bP{{\mathbf{P}}}
\def\bQ{{\mathbf{Q}}}

\def\b0{{\boldsymbol{0}}}

\newlength{\figwidth}
\newlength{\figwidthb}
\ifCLASSINFOpdf
\else
\fi

\begin{document}
\DeclareGraphicsExtensions{.eps}

\title{Optimal Power Allocation for GSVD-Based Beamforming in the MIMO Wiretap Channel\thanks{The authors are with the Dept. of Electrical Engineering and Computer Science, University of California, Irvine, CA 92697-2625, USA. e-mail:\{afakoori, swindle\}@uci.edu}}

\author{\normalsize S. Ali. A. Fakoorian, {\it Student Member, IEEE} and A. Lee Swindlehurst, {\it Fellow, IEEE}}

\maketitle

\begin{abstract}
This paper considers a multiple-input multiple-output (MIMO) Gaussian
wiretap channel model, where there exists a transmitter, a legitimate
receiver and an eavesdropper, each equipped with multiple
antennas. Perfect secrecy is achieved when the transmitter and the
legitimate receiver can communicate at some positive rate, while
ensuring that the eavesdropper gets zero bits of information. In this
paper, the perfect secrecy capacity of the multiple antenna MIMO
wiretap channel is found for aribtrary numbers of antennas under the
assumption that the transmitter performs beamforming based on the
generalized singular value decomposition (GSVD).  More precisely, the
optimal allocation of power for the GSVD-based precoder that achieves
the secrecy capacity is derived.  This solution is shown to have
several advantages over prior work that considered secrecy capacity
for the general MIMO Gaussian wiretap channel under a high SNR
assumption. Numerical results are presented to illustrate the proposed
theoretical findings.

\end{abstract}


\section{Introduction}

The broadcast nature of a wireless medium makes it very
susceptible to eavesdropping, where the transmitted message is decoded by
unintended receiver(s). Recent information-theoretic research on
secure communication has focused on enhancing security at the physical
layer. The wiretap channel, first introduced and studied by Wyner [1],
is the most basic physical layer model that captures the problem of
communication security. Wyner showed that when an eavesdropper's
channel is a degraded version of the main channel, the source and
destination can achieve a positive secrecy rate, while ensuring that
the eavesdropper gets zero bits of information. The maximum secrecy
rate from the source to the destination is defined as the secrecy
capacity. The Gaussian wiretap channel, in which the outputs at the
legitimate receiver and at the eavesdropper are corrupted by additive
white Gaussian noise (AWGN), was studied in [2].

Determining the secrecy capacity of a Gaussian wiretap channel is in
general a difficult non-convex optimization problem, and has been
addressed independently in [3-7].  Oggier and Hassibi [3] and Khisti
and Wornell [4, 5] followed an indirect approach using a Sato-like
argument and matrix analysis tools. They considered the problem of
finding the secrecy capacity of the Gaussian MIMO wiretap channel
subject to a constraint on the total average power, and a closed-form
expression for the secrecy capacity in the high signal-to-noise-ratio
(SNR) regime was obtained in [5]. However, the optimal input
covariance matrix that achieves the secrecy capacity at high SNR is
not fully characterized, especially for the case where there is
non-trivial nullspace for the channel between the transmitter and
eavesdropper. When there is such a nullspace, [5] uses a complicated
beamforming matrix to transmit two groups of information-bearing
symbols into two different subspaces, one that lies in the nullspace
of the channel matrix at the unintended receiver, and the other
orthogonal to it.  As indicated in [5], most of the transmission power
is allocated to the first subspace, and only a small fraction of the
power is allocated to the other.  Furthermore, the available power is
distributed uniformly over the dimensions of the two subspaces.
In addition to the complexity of the beamforming matrix, the other
drawback of [5] is that the precise allocation of power between the
two subspaces is not clear, nor is the sensitivity of the secrecy
capacity to this power fraction quantified.

In [6], Liu and Shamai propose a more information-theoretic approach
using the enhancement concept, originally presented by Weingarten et
al. [8], as a tool for the characterization of the MIMO Gaussian
broadcast channel capacity. Liu and Shamai have shown that an enhanced
degraded version of the channel attains the same secrecy
capacity as does a Gaussian input distribution. From the mathematical
solution in [6] it was evident that such an enhanced channel exists;
however it was not clear how to construct such a channel until the
work of [7], which provided a closed-form expression for the secrecy
capacity under a \emph{covariance matrix} power constraint.  While
this result is interesting since the expression for the secrecy
capacity is valid for all SNR scenarios, there still exists no
computable secrecy capacity expression for the MIMO Gaussian wiretap
channel under an average total power constraint.  To date, a
solution has only been obtained for the so-called MISOME case in [4],
where the transmitter and eavesdropper may have multiple antennas,
but the desired receiver has only one.

In this paper, we investigate the non-convex optimization of the
secrecy rate for the general Gaussian multiple-input multiple-output
(MIMO) wiretap channel under a constraint on the total average power,
where the number of antennas is arbitrary for both the transmitter and
the two receivers.  We focus on the case where the transmitter uses
beamforming (linear precoding) based on the generalized singular value
decomposition (GSVD), as in [5].  In particular, we obtain the optimal power
allocation that achieves the secrecy capacity for the GSVD scheme.
The resulting power allocation is significantly different in nature
than the standard water-filling solution for achieving capacity in
MIMO links without secrecy considerations.
Compared with [5], our beamforming matrix is much simpler to
compute, and more importantly, the input covariance matrix that
achieves the secrecy capacity is completely characterized in terms of
the beamforming and power allocation matrices. We note that the analysis in this paper
characterizes the secrecy capacity of the Gaussian MIMO wiretap channel with
GSVD-based beamforming for \emph{any SNR} conditions, while [5] gives
the secrecy capacity of a \emph{general} Gaussian MIMO wiretap channel
(no restriction to GSVD beamforming), but only for the \emph{high SNR}
case.

The rest of this paper is organized as follows. In the next section,
we describe the assumed mathematical model.  We then present the GSVD
method and derive the optimal power allocation that achieves the
secrecy capacity in Section~III. Finally, we demonstrate our results
by means of several numerical examples in Section~IV.

\section{System Model}  \label{sec:ach1}

Consider a multiple-antenna wiretap channel with $n_t$ transmit
antennas and $n_r$ and $n_e$ receive antennas at the legitimate
recipient and the eavesdropper, respectively:
\begin{equation}\label{eq:r1}
\by_r=\textbf{H}_r \bx+ \bz_r
\end{equation}
\begin{equation}\label{eq:r2}
\by_e=\textbf{H}_e \bx+ \bz_e
\end{equation}
where $\bx$ is a zero-mean $n_t \times 1$ transmitted signal vector,
$\bz_r\in\mathbb{C}^{n_r\times1}$ and
$\bz_e\in\mathbb{C}^{n_e\times1}$ are the additive white Gaussian
noise (AWGN) vectors at the receiver and eavesdropper, respectively,
with i.i.d. entries distributed as $\mathcal{CN}(0, 1)$.  The matrices
$\textbf{H}_r\in\mathbb{C}^{n_r\times{n_t}}$ and
$\textbf{H}_e\in\mathbb{C}^{n_e\times n_t}$ represent the channels
associated with the receiver and the eavesdropper, respectively, and
are assumed to be quasi-static flat Rayleigh fading and independent of
each other, with i.i.d. entries distributed as $\mathcal{CN}(0,
\sigma_{r}^2)$ and $\mathcal{CN}(0, \sigma_{e}^2)$.  Similar to other
papers considering the secrecy capacity of the wiretap channel, we
assume that the transmitter and both receivers are aware of the
channel state information (CSI) for both links.

In a wiretap channel, the transmitter intends to send a confidential
message $W$ to the intended receiver while keeping it as secret as
possible from the eavesdropper. The corresponding
information-theoretic secrecy constraint is given by [1, 9]:
\begin{equation}\label{eq:3}
\lim_{N\to\infty} \frac{1}{N}  I (W ; Y_e^N)=  0
\end{equation}
where $N$ is the number of channel uses for sending the message $W$,
and $I(W ; Y_e^N)$ represents the mutual information between $W$ and
$Y_e^N$. Consequently, the secrecy capacity is defined as the maximum
number of bits that can be correctly transmitted to the intended
receiver while keeping the eavesdropper uninformed. Using the
single-letter characterization of the secrecy capacity of the wiretap
channel provided by Csiszar and Korner in [9], the secrecy capacity is
the solution of the following maximization problem:
\begin{equation}\label{eq:r4}
C_{sec}=\max_{U,X} [I(U; Y_r)- I(U; Y_e)]
\end{equation}
where $X$, $Y_r$ and $Y_e$ are random variable counterparts to the
specific realizations $\bx$, $\by_r$ and $\by_e$, respectively. $U$ is
an auxiliary variable, and the maximization is over all jointly
distributed $(U,X)$ such that $U\rightarrow X \rightarrow (Y_r, Y_e)$
forms a Markov chain, while the channel input $\bx$ satisfies an
average total power constraint
\begin{equation}\label{eq:r5}
\mathrm{Tr}(E\{\bx \bx^H\})\leq p
\end{equation}
where $(.)^H$ denotes the Hermitian (i.e., conjugate) transpose, $E$
is the expectation operator, and Tr(.) is the matrix trace.

The auxiliary variable $U$ represents a precoding signal. In [6], Liu
and Shamai studied the optimization problem of (4) and showed that a
Gaussian $U = X$ is an optimal choice. In other words, Gaussian random binning
without prefix coding
is an optimal coding strategy for the MIMO Gaussian wiretap channel
[10]. Hence, a matrix characterization of the secrecy capacity is
given by
\begin{equation}\label{eq:r6}
C_{sec}=\max_{\bQ_x\succeq0} [I(X; Y_r)- I(X; Y_e)]
\end{equation}
where $\bQ_x=E\{\bx \bx^H\}$ is the input covariance matrix. The
non-convex maximization problem in (6) is considered under the power
constraint (5).

\section{Optimal Power Allocation for the GSVD-Based Gaussian MIMO Wiretap Channel}  \label{sec:ach1}

We consider the non-convex maximization problem in (6) for the case
that the transmitter performs beamforming by applying the generalized
singular value decomposition (GSVD) to the channel matrices
$\textbf{H}_r$ and $\textbf{H}_e$. Application of the GSVD technique
was first used by Khisti and Wornell (see e.g. [5]) who obtained a
closed-form expression for the secrecy capacity in the high SNR
regime. In this section, we first describe the GSVD beamforming method and next we derive 
the optimal power allocation matrix that achieves the secrecy capacity
for any SNR, and we describe some important advantages of this scheme
over what is proposed in [5].$\\$
\textbf{Definition 1} \emph{(GSVD
Transform)}: Given two matrices \textbf{H$_r$} $\in$
$\mathbb{C}^{n_r\times{n_t}}$ and \textbf{H$_e$} $\in$
$\mathbb{C}^{n_e\times n_t}$, $gsvd($\textbf{H$_r$}, \textbf{H$_e$}$)$
returns unitary matrices $\bPsi_r$ $\in$ $\mathbb{C}^{n_r\times n_r}$
and $\bPsi_e$ $\in$ $\mathbb{C}^{n_e\times n_e}$, non-negative
diagonal matrices $\bC$ and $\bD$, and a matrix $\bA$ $\in$
$\mathbb{C}^{n_t\times q}$ with $q$=min$(n_t , n_e+n_r)$, such that%
\begin{equation}\label{eq:r7}
\textbf{H$_r$}\bA = \bPsi_r\bC
\end{equation}
\begin{equation}\label{eq:r8}
\textbf{H$_e$}\bA = \bPsi_e\bD
\end{equation}
The nonzero elements of $\bC$ are in ascending order while the nonzero
elements of $\bD$ are in descending order. Moreover,
$\bC^T\bC+\bD^T\bD=\textbf{I}$.

The transmitted signal vector $\bx$ is constructed as
\begin{equation}\label{eq:r9}
\bx= \bA X_S ,\quad X_{S}\sim \mathcal{CN}(\textbf{0}, \bP)
\end{equation}
where $\bA$ is obtained from $gsvd($\textbf{H$_{r}$},
\textbf{H$_{e}$}$)$ as above, and each element of the vector $X_{S}$ represents
an independently encoded Gaussian codebook symbol that is beamformed
with the corresponding column of the $\bA$ matrix. $\bP$ is a positive
semi-definite diagonal matrix representing the power allocated by the
transmitter to the data symbols. In the following, we derive an
optimal source power allocation which achieves the secrecy capacity of
the GSVD-based MIMO Gaussian wiretap channel.  Substituting (9)
into the channel model (1)-(2) and using (7)-(8) yields
\begin{equation}\label{eq:r10}
\by_r= \bPsi_{r}\bC X_S + \bz_r
\end{equation}
\begin{equation}\label{eq:r11}
\by_e= \bPsi_{e}\bD X_S+ \bz_e
\end{equation}
Considering the above equations, the maximization problem in (6) is represented by
$$C_{sec}=\max_{\bQ_x} [I(X;Y_r)- I(X;Y_e)]=$$
$$\max_{\bP\succeq\textbf{0}, diagonal} \log|\textbf{I}+\bPsi_{r}\bC\bP\bC^{T}\bPsi_{r}^H|-\log|\textbf{I}+\bPsi_{e}\bD\bP\bD^{T}\bPsi_{e}^H|$$
\begin{equation}\label{eq:r12}
\text{subject to} \quad \text{Tr}(\bA^H \bA \bP)\leq p
\end{equation}
\textbf{Theorem 1:} Assuming that the transmitter applies the proposed
beamforming matrix $\bA$, the optimal source power allocation $\bP^*$
that achieves the secrecy capacity in problem (12) is given by
\begin{equation}\label{eq:r13}
p^*_i= \begin{cases} \max(0,\frac{-1+\sqrt{1-4c_i d_i+4(c_i-d_i)c_i d_i/(\mu a_i)}}{2c_i d_i}), & \text{if}\quad c_i > d_i\\ 0, & \mbox{otherwise}
\end{cases}
\end{equation}
where $p^*_i$, $c_i$, $d_i$ and $a_i$ are the $i$th diagonal elements
of the matrices $\bP^*$, $\bC^T\bC$, $\bD^T\bD$ and
diag$(\bA^H\bA)$, respectively. The Lagrange parameter $\mu>0$ is
chosen to satisfy the power constraint (5).$\\$
\textbf{Proof:} The optimization problem is non-convex. Our proof
technique involves applying the Karush-Kuhn-Tucker (KKT) conditions
(as necessary conditions), which help express the Lagrangian in the
form of an integral. This specific structure of the problem is then
exploited to obtain a closed-form solution for the optimal power
allocation strategy. Details can be found in the Appendix.

Eqs. (7) and (8) show that applying the GSVD transform to
\textbf{H$_r$} and \textbf{H$_e$} simultaneously diagonalizes
them. Thus, the GSVD transform creates a set of parallel independent
subchannels between the sender and the receivers, and it suffices to
use independent Gaussian codebooks across these subchannels. More
precisely, as (13) indicates, it is optimal to use only those
subchannels for which the output at the eavesdropper is a degraded
version of the output at the destination node. These subchannels
correspond to the condition $c_i> d_i$, or as shown in [5],
generalized singular values of $gsvd($\textbf{H$_r$},
\textbf{H$_e$}$)$ which are larger than 1. Clearly, if there are no
such subchannels, the achievable secrecy rate using this transmission
scheme would be zero [11].

It is interesting to note that the optimal source power allocation
(13) is different from the water-filling allocation that achieves
capacity for fading channels without the secrecy constraint. Moreover,
as we will observe in Section IV, (13) has an important role in
achieving the secrecy capacity even for moderately high SNRs. We have the
following result. $\\$ \textbf{Corollary 1:} The secrecy capacity of
the GSVD-based Gaussian MIMO wiretap channel is
\begin{equation}\label{eq:r14}
C_{sec}=\log|\textbf{I}+\bP^*\bC^T\bC|-\log|\textbf{I}+\bP^*\bD^T\bD|
\end{equation}
\textbf{Proof:} Follows directly from substituting (13) in to (12) and
by considering the fact that $\bPsi_{r}$ and $\bPsi_{e}$ are unitary
matrices.

It was shown in [5] that GSVD beamforming with uniform power
allocation is sufficient to attain the secrecy capacity of a MIMO
Gaussian wiretap channel in the high SNR regime.  However, for
the case where there is a non-trivial nullspace in the channel between the transmitter and
eavesdropper, the optimal input covariance matrix that achieves the
secrecy capacity is not fully
characterized in [5]. When there is such a nullspace, [5] uses a complicated
beamforming matrix to transmit two groups of information bearing
symbols into two different subspaces. The associated beamforming matrices are
obtained by performing an LQ decomposition on $\bA$ as well as
additional calculations which lead to Eq. (59) in [5].  The
aforementioned subspaces are identified as follows [5, Eq. (58)]:
$$\mathcal{S}_1=\text{Null}(\bH_e)\cap\text{Null}(\bH_r)^\perp$$
\begin{equation}\label{eq:r15}
\mathcal{S}_2=\text{Null}(\bH_e)^\perp\cap\text{Null}(\bH_r)^\perp
\end{equation}
where Null(.) denotes the nullspace of its matrix argument, while
$\perp$ denotes the orthogonal complement. It is important to note that
our transmission scheme in (9) and consequently our solution in Theorem 1 does not require
such calculations, and in fact yields the secrecy capacity of the Gaussian MIMO wiretap channel
with GSVD-based beamforming for any SNR under any assumptions regarding the nullspace of the
transmitter-to-eavesdropper channel.

As indicated in [5], most of the transmission power is assumed to be
allocated to the subspace $\mathcal{S}_1$ and only a small fraction
for $\mathcal{S}_2$, and the available power is distributed uniformly
over the dimensions of each of these two subspaces. The exact
allocation of power between these two subspaces is not specified in
[5], nor is the sensitivity of the secrecy capacity to the power
allocation studied. In the next section, we show that the secrecy
capacity of a MIMO Gaussian wiretap channel is in fact quite sensitive
to how power is allocated between these two subspaces, which
illustrates that using the proposed optimal power allocation is
essential.

\section{Numerical Results}  \label{sec:ach4}

In this section, we present numerical results to illustrate our
theoretical findings. In all of the following figures, the channel
matrices and background noise are modeled in the same way that we
described in Section II. In each figure, the values of $n_t$, $n_r$
and $n_e$, as well as $\sigma_r^2$, $\sigma_e^2$ and $p$, will be
depicted. In the simulations we compare the secrecy capacity obtained
by the optimal power allocation with the secrecy rate achieved by
uniform power allocation for a MIMO Gaussian wiretap channel with
GSVD-based beamforming. This comparison is performed for various
transmit powers (SNR), channel conditions and also different numbers of
antennas for the transmitter and receivers.  All displayed results are
calculated based on an average of at least 100 independent channel
realizations. In each trial, the secrecy capacity is obtained by
evaluating (14), while the secrecy rate achieved by the uniform power
allocation is obtained by using analytical results presented in
[5].

First we consider the case where the nullspace of the eavesdropper's
channel is non-trivial.  Figs. 1 and 2 investigate the allocation of
power between subspaces $\mathcal{S}_1$ and $\mathcal{S}_2$ as defined
in (15), for a case where $n_t=n_r=5, n_e=4$ and the transmit power is
$p=100$, or SNR=20 dB. The solid horizontal line illustrates the
secrecy capacity of a Gaussian MIMO wiretap channel with GSVD-based
beamforming and optimal power allocation, while the dashed curve
represents the secrecy rate achieved by uniform power allocation
versus the fraction of power used in subspace $\mathcal{S}_2$. Note
that, for this high SNR scenario, the peak of this curve is the
secrecy capacity of the \emph{general} MIMO Gaussian wiretap channel
[5].  Note also that, contrary to claims made in [5], the secrecy
rate is quite sensitive to the fraction of power allocated to the
two subspaces, and optimal performance requires a non-trivial
allocation of power to $\mathcal{S}_2$ (over 20\%).  The advantage
of the optimal GSVD-based power allocation approach is that this
imprecise distribution of power to the two subspaces is eliminated.

Fig. 3 compares the secrecy capacity achieved by the optimal power
allocation and the secrecy rate achieved by the uniform power
allocation for different transmit powers (SNRs). In this example,
there is no non-trivial null space between the transmitter and
eavesdropper. The figure shows that the difference between the optimal
and uniform power allocation is important, even at moderately high SNRs.  This is
especially true for the case where the desired receiver's channel is
of equal or better quality than the eavesdropper's channel ($\sigma_r^2\ge
\sigma_e^2$).
As the figure shows, the performance difference between
the optimal and uniform power allocation curves slowly decreases as the SNR
is increased. This is due to this fact that, as derived in [5] for the high SNR regime,
uniform power allocation is sufficient to achieve the secrecy capacity.

\section{Conclusions}

We have established the secrecy capacity for the Gaussian MIMO wiretap
channel assuming the transmitter uses GSVD-based beamforming. This
non-convex optimization problem is solved subject to an average
transmit power constraint. In particular, we have derived the optimal
power allocation for the GSVD-based beamformers that achieves the secrecy
capacity. Our numerical results demonstrate that the optimal power
allocation is necessary even for relatively high SNRs.

\appendix
We are interested in obtaining the power distribution that
achieves the secrecy capacity of the Gaussian MIMO wiretap channel in
problem (12) for the GSVD-based beamforming scheme presented in
Section III. This non-convex optimization problem is to be solved with the
average power constraint
$$\text{Tr}(E\{\bx \bx^H\})= \text{Tr}(\bA E\{X_{S} X_{S}^H\}\bA^H)$$
\begin{equation}\label{eq:r16}
=\text{Tr}(\bA \bP\bA^H)= \text{Tr}(\bA^H \bA \bP)\leq p
\end{equation}
By considering the fact that $\bPsi_{r}$ and $\bPsi_{e}$ are unitary
matrices and $\det(\textbf{I}+\bE\bF)=\det(\textbf{I}+\bF\bE)$, the
Lagrangian function $\mathcal{L}$ associated with this problem is given
by
\begin{equation}\label{eq:r17}
\mathcal{L}=\log|\textbf{I}+\bP\bC^T\bC|-\log|\textbf{I}+\bP\bD^T\bD|-\mu \text{Tr}(\bA^H\bA\bP)
\end{equation}
where $\mu>0$ is the Lagrange multiplier. Since $\bP$, $\bC^T\bC$ and
$\bD^T\bD$ are diagonal matrices, $\mathcal{L}$ can be written as
\begin{equation}\label{eq:r18}
\mathcal{L}=\sum_{i} [\log(1+p_ic_i) - \log(1+p_id_i)] -\mu \sum_{i} a_i p_i
\end{equation}
where $p_i$, $c_i$, $d_i$ and $a_i$ are the $i$th diagonal elements of
the matrices $\bP$, $\bC^T\bC$, $\bD^T\bD$ and diag$(\bA^H\bA)$,
respectively. Clearly, $p_i$, $c_i$, $d_i$ and $a_i$ all are real
non-negative numbers. Using a technique similar to that proposed in [12], the
optimal $p_i^*$ must maximize
\begin{equation}\label{eq:r19}
\mathcal{L}_i=\log(1+p_ic_i) - \log(1+p_id_i) -\mu a_i p_i=\int_0^{p_i} f_i(x)dx
\end{equation}
where $f_i(x)$ is defined as
\begin{equation}\label{eq:r20}
f_i(x)=\frac{1}{\ln 2} \left(\frac{c_i}{1+xc_i}-\frac{d_i}{1+xd_i}\right)-\mu a_i
\end{equation}
To obtain the optimal $p_i^*$ that maximizes $\mathcal{L}_i$ in (19),
we consider two cases based on the relationship between $c_i$ and
$d_i$.
\begin{enumerate}
\item $c_i\leq d_i$: In this case, (20) is always non-positive,
i.e., $f_i(x)\leq 0$. Hence, the maximum of $\mathcal{L}_i$ is
achieved by $p_i^*=0$.
\item $c_i > d_i$: In this case, since $f_i(x)$ is a decreasing function
for $x\geq 0$, the optimal $p_i^*$ that maximizes $\mathcal{L}_i$
depends on the value of the largest root of $f_i(x)=0$. Let $x_{Li}$
denote the largest root of $f_i(x)=0$, i.e.,
$$x_{Li}=\frac{-1+\sqrt{1-4c_i d_i+4(c_i-d_i)c_i d_i/(\mu a_i)}}{2c_i
d_i}$$ where we have used the fact that
$\bC^T\bC+\bD^T\bD=\textbf{I}$, or equivalently
$c_i+d_i=1$.\footnote{It is easy to verify that $\mu>0$ guarantees
$1-4c_i d_i+4(c_i-d_i)c_i d_i/(\mu a_i)\ge 0$ for the case $c_i >
d_i$. To do so, it is sufficient to prove that $1-4c_i d_i\ge 0$. We
have:$$c_i+d_i=1\rightarrow1-4c_i d_i= 1-4
(1-d_i)d_i=1-4d_i+4d_i^2=(1-2d_i)^2\ge 0$$ } If $x_{Li}$ is positive
then the maximum of $\mathcal{L}_i$ is achieved by $p_i^*=x_{Li}$,
otherwise $p_i^*=0$.
\end{enumerate}
Combining cases 1 and 2, we obtain the desired result as
\begin{equation}\label{eq:r21}
p_i^*= \begin{cases} \max(0,\frac{-1+\sqrt{1-4c_i d_i+4(c_i-d_i)c_i d_i/(\mu a_i)}}{2c_i d_i}), & \text{if}\quad c_i > d_i\\ 0, & \mbox{otherwise}
\end{cases}
\end{equation}
Finally, the Lagrange parameter $\mu>0$ is chosen to satisfy the power constraint (16).

\begin{figure}[!t]
\begin{center}
\includegraphics[width=5in,height=5in]{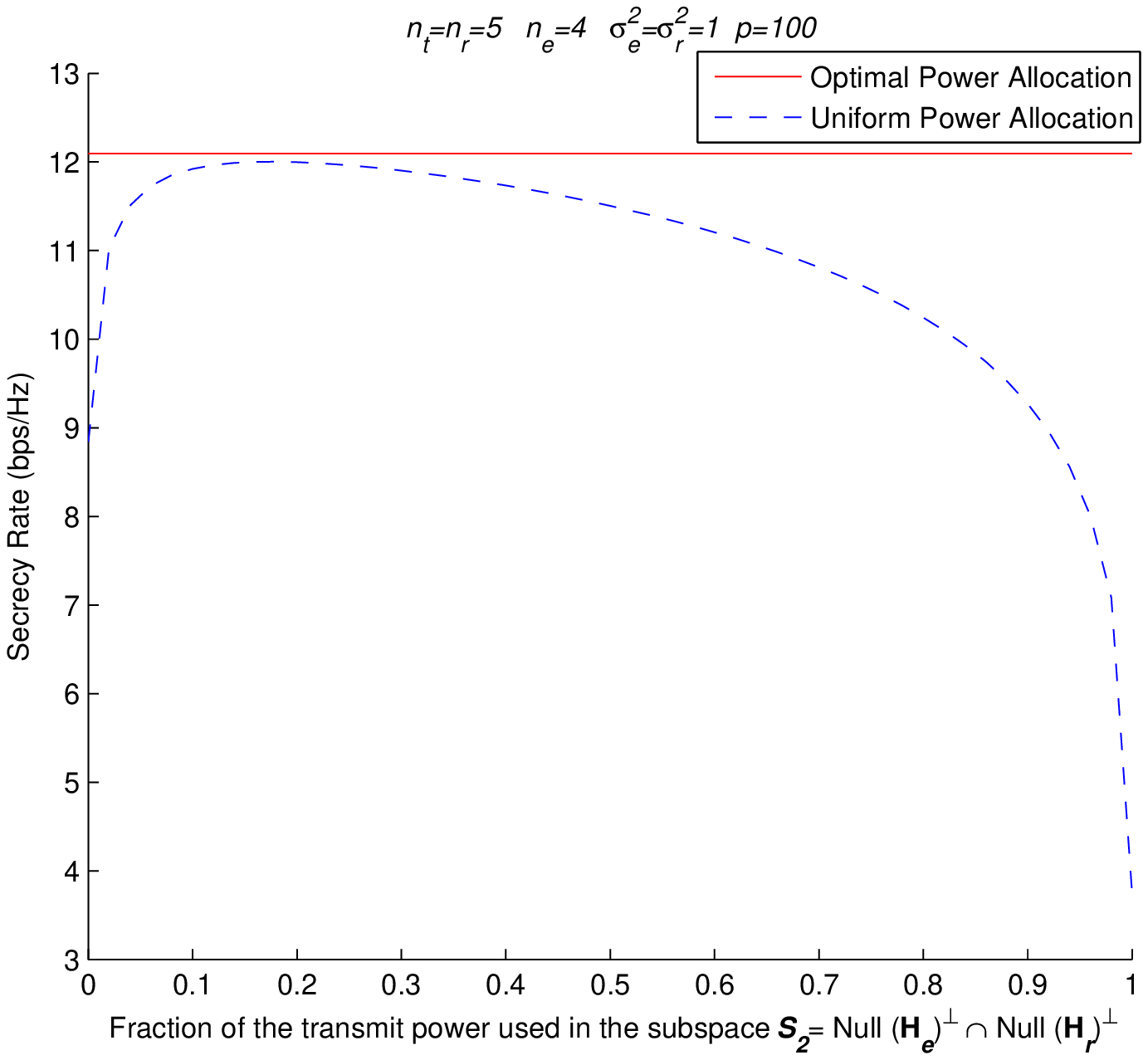}
\end{center}
\caption{Comparison of secrecy capacity for optimal power allocation
with secrecy rate for uniform power allocation at high SNR in a
low interference scenario.}
\label{fig_sim}
\end{figure}

\begin{figure}[!t]
\begin{center}
\includegraphics[width=5in,height=5in]{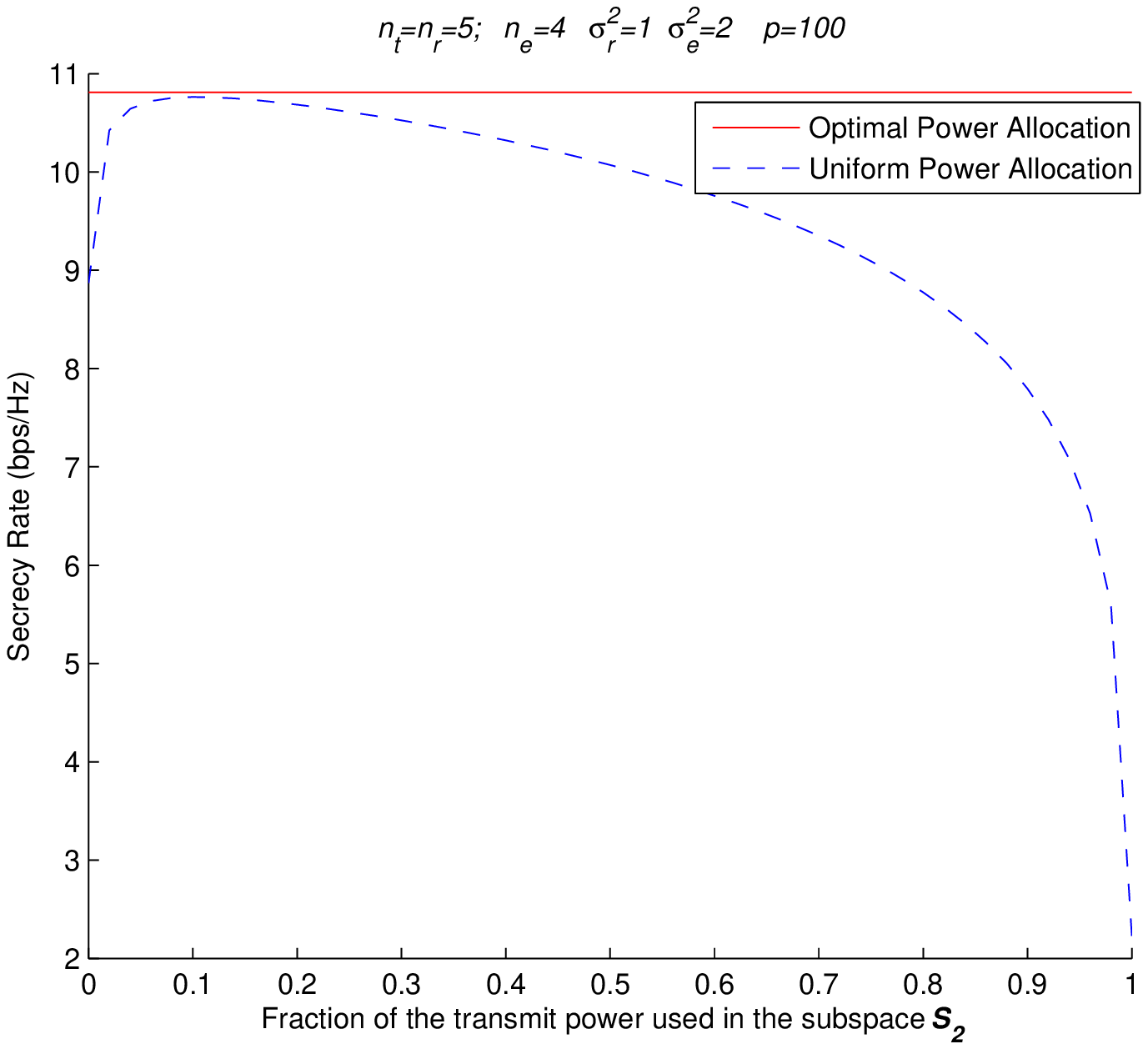}
\end{center}
\caption{Comparison of secrecy capacity for optimal power allocation
with secrecy rate for uniform power allocation at high SNR in a
high interference scenario.}
\label{fig_sim}
\end{figure}

\begin{figure}[!t]
\begin{center}
\includegraphics[width=5in,height=5in]{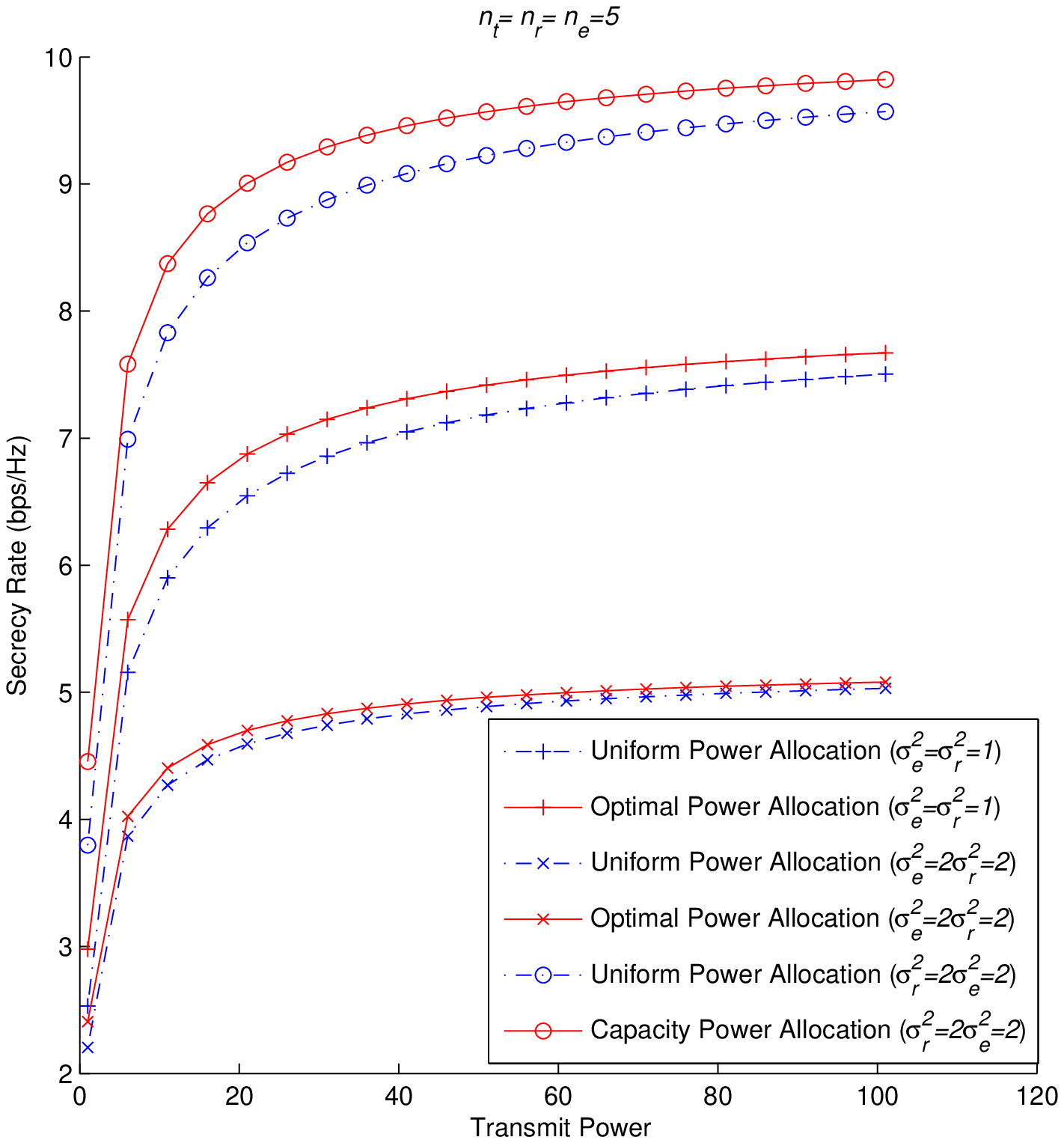}
\end{center}
\caption{Comparison of secrecy capacity by optimal power allocation with secrecy rate by uniform power allocation under different SNR and interference situations.}
\label{fig_sim}
\end{figure}


\begin{thebibliography}{1}

\bibitem{1}
A.~Wyner, ``The wire-tap channel,'' \emph{Bell. Syst. Tech. J.}, vol. 54, no. 8, pp. 1355-1387, Jan. 1975.
\bibitem{2}
S.~K.~Leung-Yan-Cheong and M.~E.~Hellman, ``The Gaussian wire-tap channel,'' \emph{IEEE Trans. Inf. Theory}, vol. 24, pp. 451-456, Jul. 1978.
\bibitem{3}
F.~Oggier and B.~Hassibi, ``The secrecy capacity of the MIMO wiretap channel,'' in \emph{Proc. IEEE Int. Symp. Information Theory} Toronto, ON, Canada, Jul. 2008, pp. 524-528.
\bibitem{4}
A.~Khisti and G.~Wornell, ``Secure transmission with multiple antennas I: The MISOME
wiretap channel,'' to appear, \emph{IEEE Trans. Inf. Theory}, 2010. Available at: http://arxiv.org/abs/0708.4219
\bibitem{5}
A.~Khisti and G.~Wornell, ``Secure transmission with multiple antennas II: The MIMOME
wiretap channel,'' to appear, \emph{IEEE Trans. Inf. Theory}, 2010. Available at: http://allegro.mit.edu/pubs/posted/journal/2008-khisti-wornell-it.pdf
\bibitem{6}
T.~Liu and S.~Shamai (Shitz), ``A note on secrecy capacity
of the multi-antenna wiretap channel,'' \emph{IEEE Trans. Inf. Theory}, vol. 55, no. 6, pp. 2547-2553, 2009.
\bibitem{7}
R.~Bustin, R.~Liu, H.~V.~Poor, and S.~Shamai (Shitz), ``A MMSE approach to the secrecy capacity of the MIMO Gaussian wiretap channel,'' \emph{EURASIP Journal on Wireless Communications and Networking}, vol. 2009, Article ID 370970, 8 pages, 2009.
\bibitem{8}
H.~Weingarten, Y.~Steinberg, and S.~Shamai (Shitz), ``The capacity region of the Gaussian multiple-input multipleoutput broadcast channel,'' \emph{IEEE Trans. Inf. Theory}, vol. 52, no. 9, pp. 3936-3964, 2006
\bibitem{9}
I.~Csiszár and J.~Körner, ``Broadcast channels with confidential messages,'' \emph{IEEE Trans. Inf. Theory}, vol. IT-24, no. 3, pp. 339-348, May 1978.
\bibitem{10}
Ruoheng Liu, Tie Liu, H.~Vincent Poor, and Shlomo Shamai (Shitz), ``Multiple-Input Multiple-Output Gaussian Broadcast Channels with Confidential Messages,'' \emph{IEEE Trans. Inf. Theory}, to appear.
\bibitem{11}
Yingbin Liang, H.~Vincent Poor, Shlomo Shamai (Shitz), ``Secure Communication over Fading Channels,'' \emph{IEEE Trans. Inf. Theory}, vol. 54, no. 6, pp. 32470-2492, Jun. 2008.
\bibitem{12}
Ruoheng Liu, Yingbin Liang and H.~Vincent Poor, ``Fading Cognitive Multiple-Access Channels With Confidential Messages,'' \emph{IEEE Trans. Inf. Theory}, submitted, Dec. 2009.
\end{thebibliography}
\end{document}